\documentclass{PoS}
\usepackage{subfig}
\usepackage{url}
\usepackage{array}
\usepackage{makecell}

\usepackage{lineno}

\title{The Transient program of the Cherenkov Telescope Array}

\ShortTitle{The CTA Transient program}

\author{\speaker{Fabian Sch\"ussler}\\
        IRFU, CEA, Universit\'e Paris-Saclay, F-91191 Gif-sur-Yvette, France\\
        E-mail: \email{fabian.schussler@cea.fr}}
\author{for the CTA consortium\footnote{for collaboration list see PoS(ICRC2019)1177}}        

\abstract{
The Cherenkov Telescope Array (CTA) is the next generation high-energy gamma-ray observatory. It will improve the sensitivity of current instruments up to an order of magnitude, while providing energy coverage for photons from 20 GeV to at least 300 TeV to reach high redshifts and extreme accelerators and will give access to the shortest time-scale phenomena. CTA is thus a uniquely powerful instrument for the exploration of the violent and variable universe. 

The ability to probe short timescales at the highest energies will allow CTA to explore the connection between accretion and ejection phenomena surrounding compact objects, investigate the processes occurring in relativistic outflows, and open up significant phase space for serendipitous discoveries. Aiming at playing a central role in the era of multi-messenger astrophysics, the CTA Transient program includes follow-up observations of a broad range of multi-wavelength and multi-messenger alerts, ranging from Galactic compact object binary systems to novel phenomena like Fast Radio Bursts. A promising case is that of gamma-ray bursts (GRBs), where CTA will for the first time enable high-statistics measurements above $\sim$ 10 GeV, probing new spectral components and shedding light on the physical processes at work in these systems. Dedicated programs searching for very-high-energy (VHE) gamma-ray counterparts to gravitational waves and high-energy neutrinos complete the CTA transients program.

This contribution will introduce and outline the CTA Transients program. We will provide an overview of the various science topics and discuss the links to multi-messenger and multi-wavelength observations.
}

\FullConference{36th International Cosmic Ray Conference -ICRC2019-\\
		July 24th - August 1st, 2019\\
		Madison, WI, U.S.A.}

\begin{document}

\section{Introduction}
\begin{figure}[!t]
  \centering
  \subfloat[CTA telescopes]{\label{fig:CTA-Telescopes}%
\includegraphics[width=0.48\textwidth]{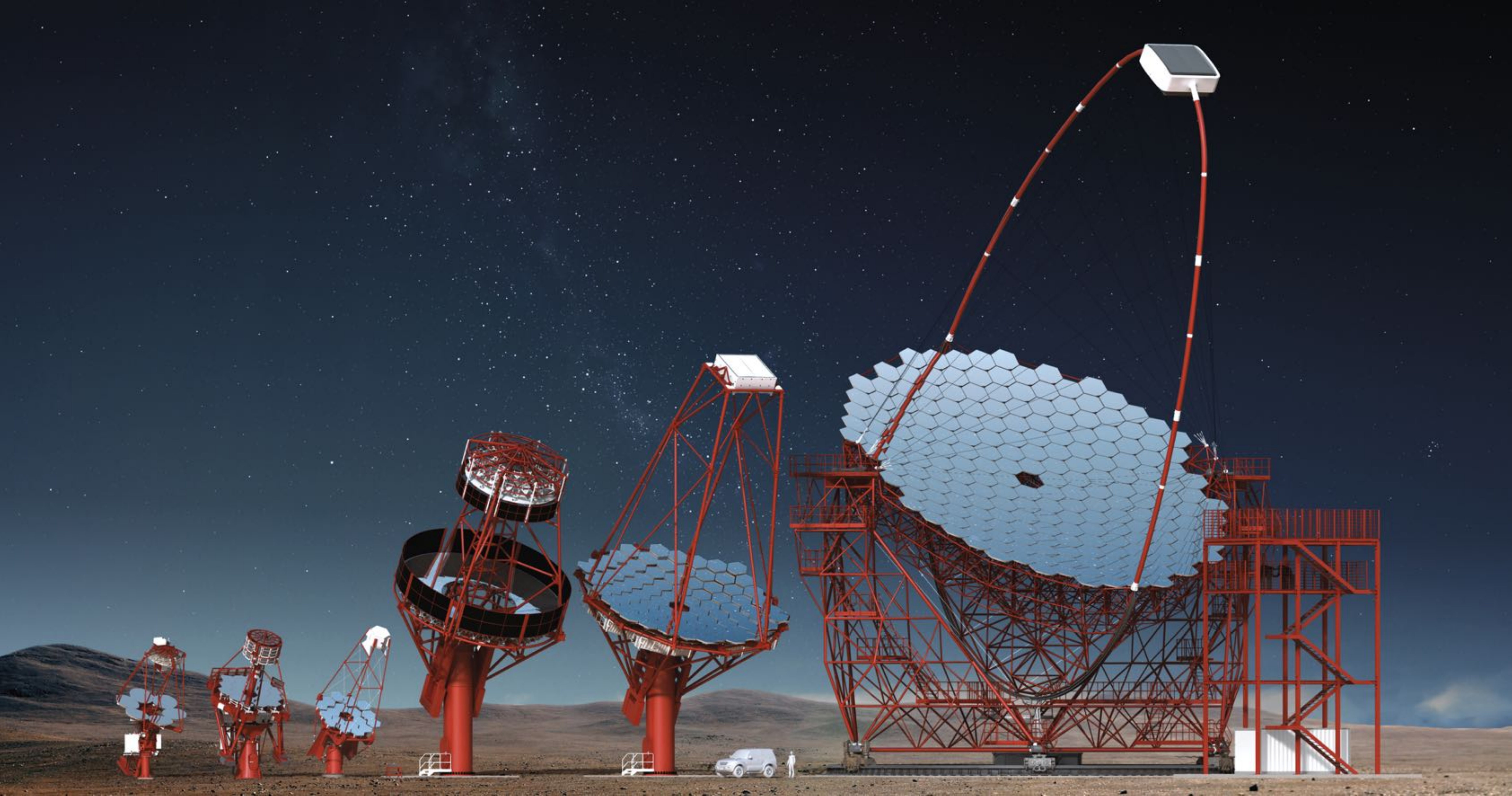}
}%
  \quad%
  \subfloat[CTA transient sensitivity]{\label{fig:CTA:sens}%
  \includegraphics[width=0.43\textwidth]{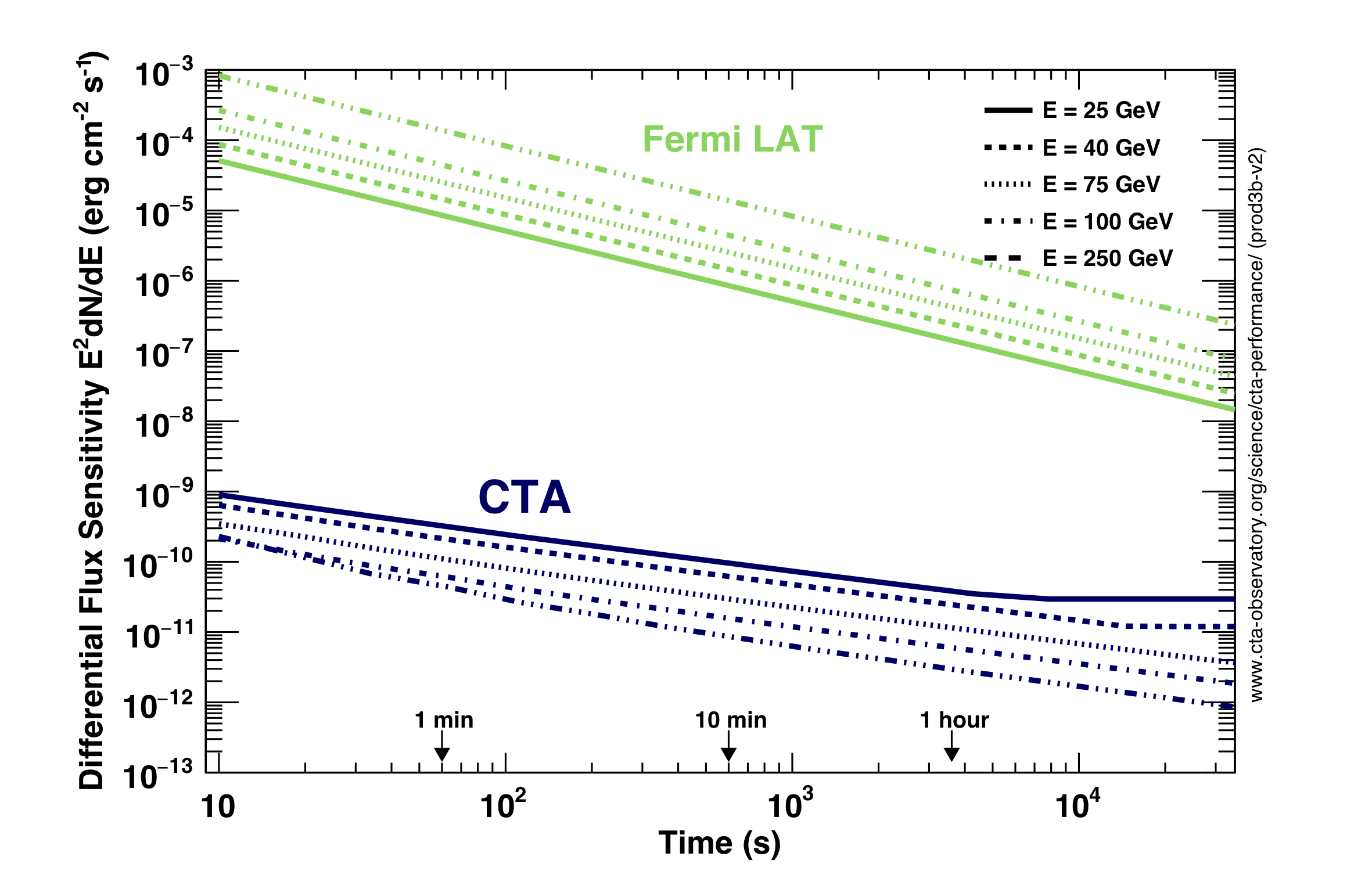}
}%
  \caption[CTA]{Left: Schematic view of the different telescope types that will form the Cherenkov Telescope Array observatory. Right: CTA sensitivity to transient phenomena (from \protect\url{cta-observatory.org}~\cite{CTAsensICRC2019}).}\label{fig:CTA}
\end{figure}

After more than a decade of preparation of the next generation, IACT-based, high-energy gamma-ray observatory, the Cherenkov Telescope Array (CTA) is currently entering its construction phase. After completion, currently expected in the mid-2020s, the CTA observatory will operate more than 100 IACTs on two sites, one in each hemisphere. The northern hemisphere array, located at the Observatorio del Roque de los Muchachos Roches on the island of La Palma, will be more limited in size and will focus on the low- and mid-energy range from 20 GeV to 20 TeV. The southern hemisphere array, located in Chile near the Paranal site of the European Southern Observatory, will span the entire energy range, covering gamma-ray energies from 20 GeV to 300 TeV. Three classes of telescope will be distributed across these two sites based on their sensitivity: Small-Sized Telescopes (SST), Medium-Sized Telescopes (MST), and Large-Sized Telescopes (LST), cf. Figure~\ref{fig:CTA}. The SST array is optimized for the highest energies and will be exclusively deployed in the southern site providing access to the Galactic plane and its wealth of high-energy sources. The MSTs and LSTs will be installed on both sites. Commissioning of the first LST started in late 2018.

In addition to a low energy threshold, the LSTs are also able to slew within less than 30 seconds to any position in the sky. They therefore make CTA well-suited to react to observations and alerts from the multi-wavelength (MWL) and multi-messenger community. Combining fast reaction, (comparably) low energy threshold and high sensitivity (cf. Figure~\ref{fig:CTA:sens}), CTA will be able to study a variety of high-energy transient phenomena in unprecedented detail. Preparations for these observations are currently ongoing within the CTA consortium and the CTA observatory. A selection of ongoing activities is outlined in the following.

\subsection{Preparations for the CTA transient program}\label{sec:CTA:transients}
Preparations for the CTA science operations started with the definition of the Key Science Projects (KSPs,~\cite{CTAScience}). Based on and extending the current state-of-art multi-wavelength and multi-messenger follow-up programs by the current IACT experiments, the Transient KSP will comprise observations of several interconnected source classes. The different sources and their priority, as well as a first estimate for the observation time that they will be allocated by the CTA observatory, is given in Table~\ref{table:KSP}. With the start of science observations of the CTA observatory still some years in the future, and given the rapid evolution of the field of multi-messenger transients, one should note that these numbers can only be viewed as preliminary estimates. Significant adjustments and updates are currently being discussed.

\noindent

\begin{table*}[!h]
\begin{center}
\caption[]{Summary of proposed observations within the CTA Transients Key Science Project~\cite{CTAScience}}
\label{table:KSP}
\smallskip
\begin{tabular}{lllll}
\hline \hline
 \multicolumn{5}{c}{Observation times (h~yr$^{-1}$~site$^{-1}$)} \\
Priority & Target class  & Early phase & Years 1--2 & Years 3--10  \\
\hline
1 & GW transients & 20 & 5 & 5 \\
2 & HE neutrino transients                        & 20                    & 5         & 5          \\
3 & Serendipitous detections               & 100                   & 25         & 25           \\
4 & GRBs                                     & 50                    & 50         & 50           \\
5 & MWL transients & 50 & 10 & 10 \\
6 & Galactic                                 & 150                   & 30          & 0           \\

\hline
Prel. total observation time & &390                   & 125        & 95        \\
\hline
\end{tabular}
\end{center}
\end{table*}

After the definition of the KSPs, the details of the future operation and observations with CTA were defined as {\it use cases}. These were produced at different levels, ranging from the individual subsystems to the full operation of the observatory and then combined in a high-level layer, the {\it Top Level Use Cases}~\cite{Bulgarelli:2016hrv}. Partially based on the experience with current IACTs, detailed observation scenarios ranging from the reception of the alerts, their analysis, the reaction of the CTA infrastructure, the data taking, and the final physics analyses have been outlined. 

Reacting to recent observational results like the MWL signatures of binary neutron star mergers (cf. Section~\ref{sec:GWs}), the detection of a flaring blazar possibly associated with a high-energy neutrino (cf. Section~\ref{sec:neutrinos}) and the first detections of GRBs by ground-based IACTs (cf. Section~\ref{sec:GRBs}), the outlined scenarios are constantly being revised and improved.

\subsection{The multi-wavelength and multi-messenger context}
CTA will operate in a rich MWL and multi-messenger environment. To name just a few: the Virgo and LIGO interferometers are being improved continuously and will pave the way towards third generation instruments like the Einstein Telescope and the Cosmic Explorer. The IceCube neutrino telescope is continuing operations while IceCube-Gen2, GVD and KM3NeT are being prepared and built. In the X-ray domain Spectr-RG, including the eROSITA instrument, is about to be launched. SVOM, being dedicated to transient phenomena, and later the multi-purpose ATHENA satellites are being constructed. The pathfinders of the Square Kilometer Array (SKA) radio observatory have started operations and the full SKA is approaching fast. In the optical domain transient factories, like ZTF and soon LSST are providing a wealth of data. At the very-high energies, the HAWC observatory has produced a number of novel results at a high rate, while the next-generation observatory LHAASO is being constructed and commissioned. 

Agreements for joint science programs between these observatories are being defined on a case-by-case basis depending on the individual needs of each topic and analysis. This definition is currently in progress within the CTA consortium. A summary of the current status is given in~\cite{MWLsynergies_ICRC2019}.

\subsection{Detecting transients in real-time}
\begin{figure}[!t]
  \centering
  \subfloat[CTA data flow]{\label{fig:dataflow}%
  \includegraphics[width=0.5\textwidth]{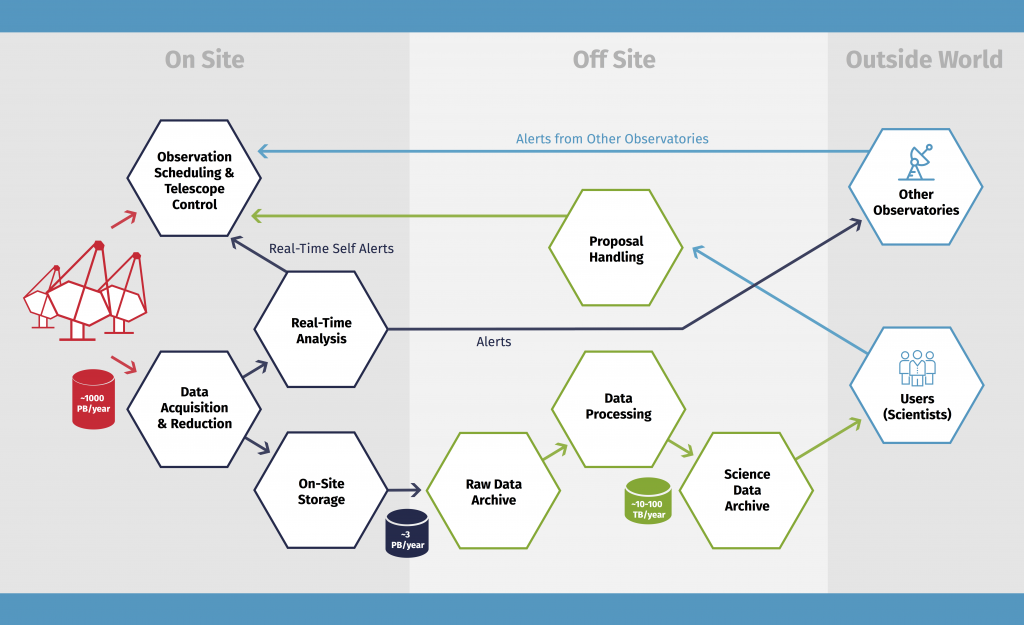}
}%
  \quad%
  \subfloat[POSyTIVE: GRB population study]{\label{fig:CTA:GRBs}%
  \includegraphics[width=0.46\textwidth]{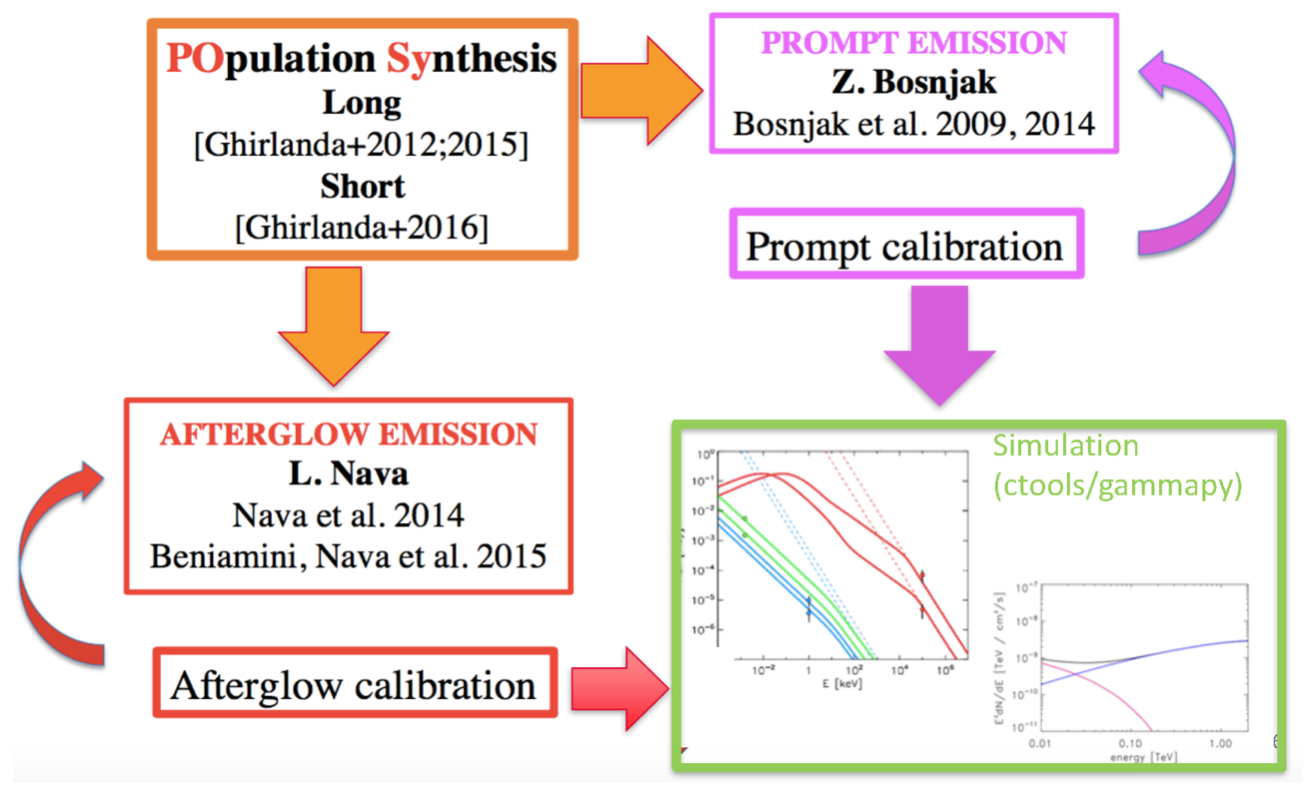}
}%
  \caption[data flow]{Left: The flow of data obtained by CTA illustrating the connection to other observatories and the central role of the RTA (from \protect\url{cta-observatory.org}). Right: Schematic overview of the main components of the POSyTIVE project, studying the potential of GRB detections with CTA. See~\cite{CTAGRBS_ICRC2019} for details.}
\end{figure}

A major advantage of CTA observations and searches for transient phenomena will be its capability to analyze the data while it is being recorded. This {\it real-time analysis} (RTA,~\cite{2015ICRC...34..763B}) will scan the incoming data for previously unknown sources or flux variations from known sources with a latency of less than 30 seconds. Alerts will be emitted both internally, e.g. to adapt the CTA observation schedule and externally to the community, e.g. to allow for rapid MWL/MM follow-up observations. A schematic view of the foreseen data flow is given in Fig.~\ref{fig:dataflow}.

RTA alerts will be used internally to trigger extended CTA follow-up observations in a large variety of scenarios. These comprise the detection of transient phenomena in targeted observations like the monitoring of AGN and follow-up observations of external alerts (e.g. GRBs, GWs, high-energy neutrinos, etc.). As the RTA analysis is constantly scanning the comparably large CTA field-of-view, serendipitous detections of (possibly novel) transient phenomena are possible as well. The large surveys of both the Galactic Plane as well as a large portion of the extragalactic sky that are foreseen with CTA~\cite{CTAScience} will provide opportunities for such serendipitous discoveries. These may establish new VHE emitters and source classes as happened recently in the radio domain with Fast Radio Bursts (FRBs).

FRBs are one of the major astronomical mysteries that have emerged in the last decade and are now established as new class of radio emitters. While FRBs release enormous amounts of energy in the radio domain, it remains the only band they have been conclusively detected in so far. While their nature remains elusive, their origins are potentially similar to other transients seen in the X-ray and gamma-ray bands, such as short and/or long GRBs~\cite{2014ApJ...780L..21Z}. Several FRB models have also specifically predicted flares in the TeV band (e.g., \cite{Lyubarsky:2014,2016MNRAS.461.1498M}) and proposed follow-ups of FRBs at very high energies. First searches for VHE afterglows of FRBs have been performed by the H.E.S.S. collaboration~\cite{2017A&A...597A.115H}. Following the detection of repeating bursts from FRB121102, coordinated MWL campaigns became possible. First results of these campaigns have been presented by the VERITAS~\cite{2017ICRC...35..621B} and MAGIC collaborations~\cite{2018MNRAS.481.2479M}. CTA will continue these searches.

Another novel gamma-ray emitting source class has been established by \textit{Fermi}-LAT after the detection of several Galactic novae~\cite{2014Sci...345..554A, 2018A&A...609A.120F}, thermonuclear explosions on the surface of a white dwarf which is accreting matter from a stellar companion. While both leptonic and hadronic models are able to explain the observed emission, the details of the particle acceleration processes reaching the observed energies remain unresolved. While novae haven't been detected by IACTs yet~\cite{2015A&A...582A..67A}, deep observations with CTA may help to provide missing pieces of this puzzle.

Microquasars~\cite{1998Natur.392..673M}, stellar-mass black holes that appear in many aspects like scaled down versions of quasars, have long been suspected to accelerate particles and thus emit VHE gamma rays. The recent discovery of persistent VHE radiation associated to the jets of SS433 by the HAWC collaboration~\cite{2018Natur.562...82A} opens the way for a new understanding of the connection between the accretion of matter onto black holes and the phenomenology of relativistic jets. The extensive and deep survey of the Galactic Plane with CTA~\cite{CTAScience} will make it possible to cover a significant fraction of the Galactic microquasars (and many other Galactic sources potentially showing transient emission) and may provide crucial observational input to these questions.

\noindent

\section{Gamma-ray Bursts}\label{sec:GRBs}
After a decade of intense preparations and searches, gamma-ray bursts have finally been detected with ground-based imaging atmospheric Cherenkov telescopes. Following first hints from GRB 160821B~\cite{MAGIC_GRB160821B_MG15}, the MAGIC experiment observed high-energy emission above 300 GeV in the early afterglow phase of GRB190114C~\cite{MAGIC_GRB190114_ATEL}, while H.E.S.S. was able to detect gamma-rays above 100 GeV in the late afterglow of GRB180720B~\cite{GRB180720B-HESS-CTAsymposium}. With its superior performance, CTA will be able to characterize the high-energy emission of gamma-ray bursts (GRBs) in detail and will thus be able to probe radiation and particle acceleration mechanisms within GRB jets. In preparation for this observation program, the CTA consortium is developing the POSyTIVE framework. With this {\it POpulation Synthesis Theory Integrated code for Very high energy Emission}, a bottom-up population model of GRBs is constructed using a minimal set of intrinsic properties and assumptions (e.g. the distribution of peak energies and redshifts as well as correlations between the peak energy and luminosity). The derived GRB population, comprising both long and short GRBs, is calibrated against the entire 40-year dataset of multi-wavelength GRB observations before both the prompt and the afterglow emission are simulated following state of the art models accounting for synchrotron and Inverse Compton components. Subsequent detailed simulation studies will make it possible to estimate the CTA GRB detection rate and provide insights into the expected wealth of observational data such as high resolution lightcurves and time-resolved energy spectra. A schematic overview of the POSyTIVE project is given in Figure~\ref{fig:CTA:GRBs} and details are given in~\cite{CTAGRBS_ICRC2019}.

\section{Gravitational waves}\label{sec:GWs}
\begin{figure}[!t]
  \centering
  \subfloat[Scheme of CTA reaction to GWs]{\label{fig:CTA:GWreception}%
  \includegraphics[width=0.53\textwidth]{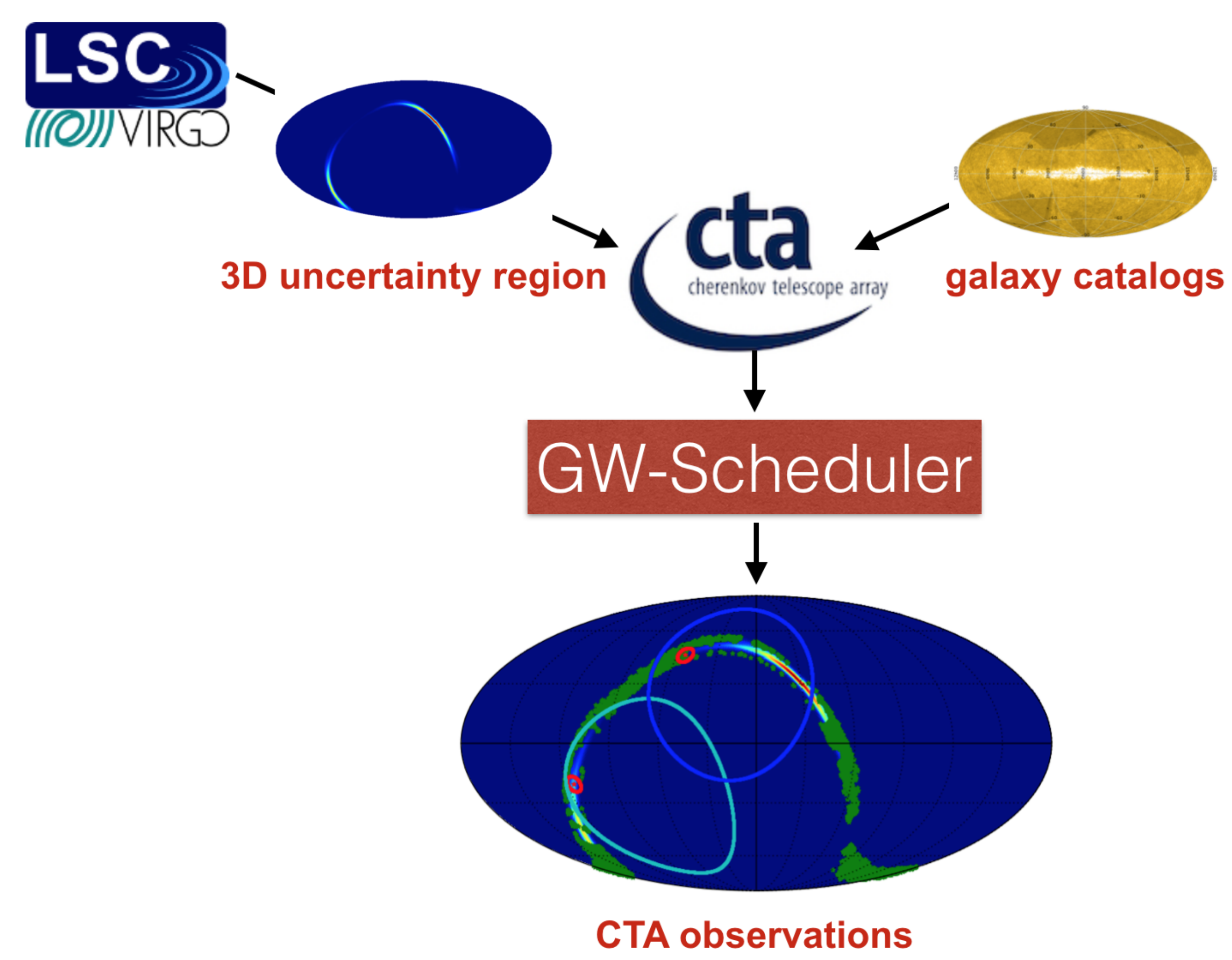}
}%
  \quad%
  \subfloat[Simulated CTA follow-up of GW170817]{\label{fig:CTA:GW170817}%
  \includegraphics[width=0.43\textwidth]{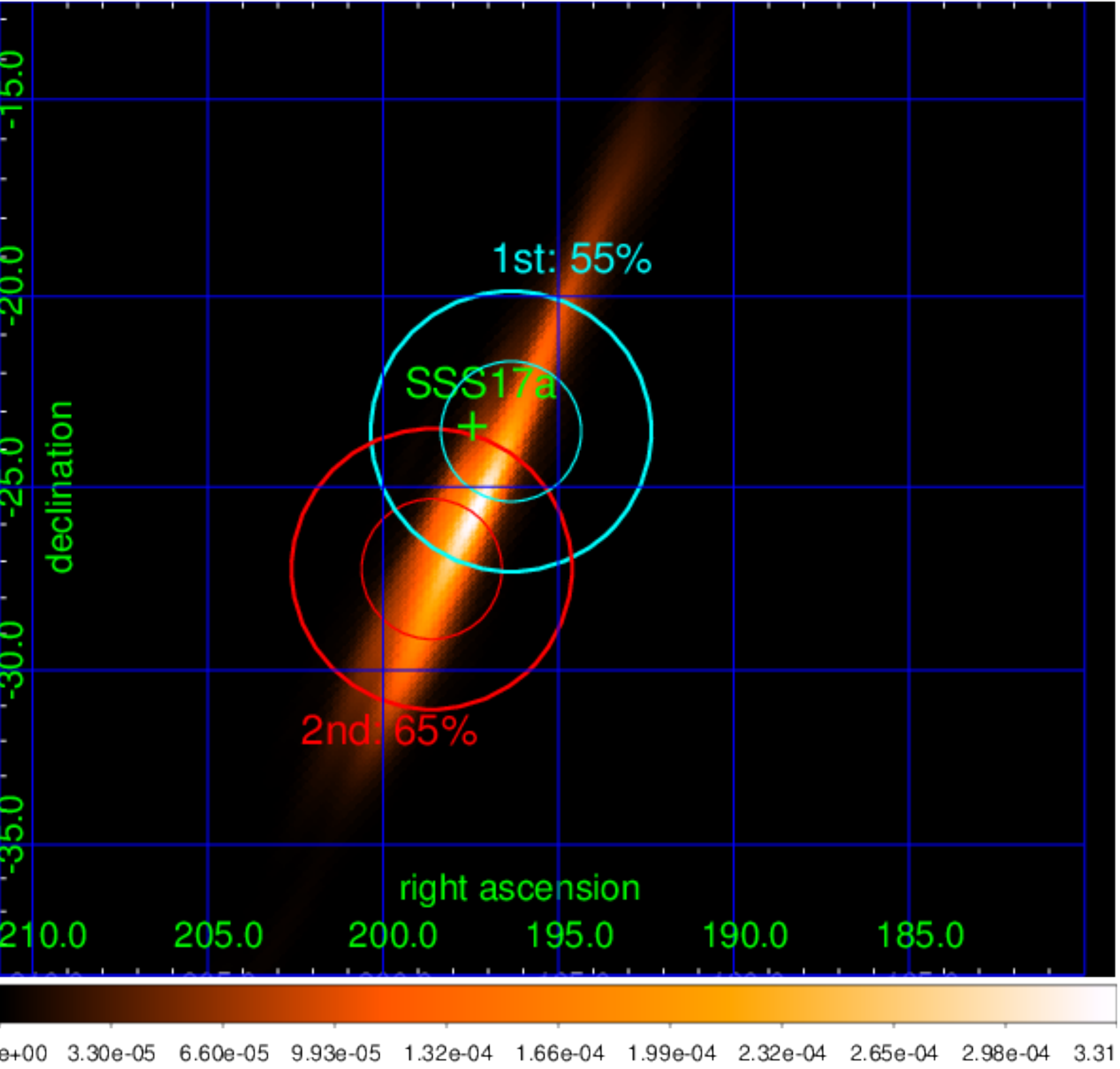}
}%
  \caption[CTA: GW follow-up]{Left: The workflow of the reaction of CTA to the detection of a gravitational wave as outlined in the CTA Key Science Projects and the CTA Science Top Level Use Cases. Right: Simulated response of CTA to the binary neutron star merger GW170817, showing the two individual observations that would be necessary to cover the majority of the GW uncertainty region. The inner (outer) lines illustrate the FoV of the LSTs (SSTs) and the numbers indicate the coverage of the GW region by pointings of the full CTA array. Figure taken from~\cite{CTA_Transients_TeVPA2018}.}
\end{figure}

The coincidence of GRB170817A and the gravitational signal GW170817 established the link between short GRBs and gravitational waves (GWs). The detection of the potential associated VHE gamma-ray emission is challenging due to the relatively large localisation uncertainties provided by the current and next generation GW interferometers~\cite{2018LRR....21....3A}. To tackle the associated difficulties, dedicated GW alert reception and scheduling algorithms are being developed within the CTA consortium. A schematic view of the planned treatment of gravitational wave alerts by CTA is given in Figure~\ref{fig:CTA:GWreception}. As can be seen, the crucial component in this scheme is the {\it CTA GW-Scheduler}, which is the tool that combines all available information (e.g. the GW uncertainty map, galaxy catalog(s), status information of the CTA array, etc.) and derives an optimal observation strategy and schedule. A first version of this tool has been implemented and is being used to study the potential performance of the CTA GW follow-up. An illustration is given in Figure~\ref{fig:CTA:GW170817}, where the simulated CTA response to the binary neutron star merger event GW170817 is given. The figure shows the (only) two individual observations that would be necessary to cover the majority of the GW uncertainty region (individual coverages assuming the 8deg FoV of the CTA SSTs are given as percentage). Similar to the H.E.S.S. observations of GW170817~\cite{2017ApJ...850L..22A} the first pointing would already contain the (\textit{a} \textit{priori} unknown) direction of the merger event.

\noindent

Further more detailed studies of the CTA performance of these observations are currently being done using an extensive set of simulated neutron star mergers~\cite{PatricelliDB}. The expected associated high-energy electromagnetic emission is estimated using the same prompt and afterglow models mentioned in Section~\ref{sec:GRBs}. The GW scheduling simulations take into account all known observational effects like the visibility from both CTA sites as well as observational constraints like dark-time and moon-distance on an event-by-event basis. They will make it possible to estimate the number of successful CTA detections of high-energy counterparts of GW events and provide first insights into the potential of these observations in the broad multi-wavelength and multi-messenger context expected for the next decade. Further details are given in~\cite{CTAGWs_ICRC2019}.

\section{High-energy neutrinos}\label{sec:neutrinos}
While the detection of an astrophysical flux of high-energy neutrinos by IceCube provides first clues towards the long-sought origins of high-energy cosmic rays, no significant small scale excess pointing towards individual sources has been detected by high-energy neutrino telescopes so far. This may hint towards a large population of faint, stable sources or to transient phenomena as origins of the detected neutrino flux. The current most compelling evidence for an individual neutrino point source is the recent observation of the flaring gamma-ray blazar TXS 0506+056 in coincidence with the high energy neutrino IC-170922A detected by IceCube~\cite{2018Sci...361.1378I}. The extensive MWL observation campaign that allowed to obtain this first evidence is the result of significant efforts installing Neutrino Target of Opportunity (NToO) programs with all major neutrino telescopes and the participation of all currently operating IACTs.

As a extension of these programs, the CTA NToO program is currently being defined. Relying on a variety of alerts from neutrino observatories, we study and optimize the chances for the detection of VHE gamma-ray counterparts to neutrino emitters with CTA. The starting point for these studies is the FIRESONG simulation framework, which provides simulated neutrino source populations matching the observed diffuse flux. The large parameter spaces (e.g. densities of sources and their cosmological evolution, different transient emission times, etc.) of different scenarios are being scanned. Relying on typical $pp$ (and later $p\gamma$) models, the gamma-ray emission associated with the simulated neutrino sources is estimated and used as input to detailed CTA detector simulations. Several CTA array layouts and instrument response functions are tested in order to derive optimal follow-up strategies and finally estimate the potential science reach of the NToO program for CTA. Further details are given in~\cite{CTANeutrinos_ICRC2019}.

\section{Outlook}
High-energy astrophysics has seen significant changes in the last decade. To name just a few: Gravitational Waves were detected and could be linked to Gamma-Ray Bursts, high-energy neutrinos of astrophysical origin were observed and could be linked (at least one of them) to a flaring blazar, based on scans of the Galactic Plane a large number of new VHE gamma-ray emitting sources were identified and finally VHE gamma-rays from GRBs were observed with ground-based IACTs. These and many more discoveries have opened new windows to the high-energy universe and thus promise even more exciting observations and discoveries in the years to come. As outlined in this contribution, preparation for these opportunities are well on track within the CTA consortium.

\section{Acknowledgements}
This work was conducted in the context of the CTA Transients Working Group. We gratefully acknowledge financial support from the agencies and organizations listed here: \url{http://www.cta-observatory.org/consortium\_acknowledgments}.


\bibliographystyle{JHEP}
\bibliography{ref.bib}

\end{document}